\documentclass[conference]{IEEEtran}
\IEEEoverridecommandlockouts
\usepackage{cite}
\usepackage{amsmath,amssymb,amsfonts}
\usepackage{algorithmic}
\usepackage{graphicx}
\usepackage{textcomp}
\usepackage{xcolor}
\usepackage{soul}
\usepackage{algorithm}
\usepackage{algorithmic}
\usepackage{arydshln}
\usepackage{makecell}
\usepackage{svg}
\usepackage[left=0.65in, right=0.65in, bottom=1.05in, top=0.72in]{geometry}

\usepackage[caption=false]{subfig}
\def\BibTeX{{\rm B\kern-.05em{\sc i\kern-.025em b}\kern-.08em
    T\kern-.1667em\lower.7ex\hbox{E}\kern-.125emX}}

\renewcommand{\baselinestretch}{0.88}

\begin{document}

\title{Dynamic Online Modulation Recognition using Incremental Learning 
\thanks{This material is based upon work supported by the Air Force Office of Scientific Research under award number FA9550-20-1-0090 and the National Science Foundation under Grant Numbers  CNS- 2318726, and CNS-2232048.} \thanks{DISTRIBUTION STATEMENT A: Approved for Public Release; distribution unlimited AFRL-2023-5606 on Nov. 06, 2023.}}

\author{
	\IEEEauthorblockN{
	Ali Owfi\IEEEauthorrefmark{1}, 
        Ali Abbasi\IEEEauthorrefmark{2},
        Fatemeh Afghah\IEEEauthorrefmark{1},
        Jonathan Ashdown \IEEEauthorrefmark{3},
        Kurt Turck \IEEEauthorrefmark{3}}

    \IEEEauthorblockA{\IEEEauthorrefmark{1}Holcombe Department of Electrical and Computer Engineering, Clemson University, Clemson, SC, USA \\
        Emails: \{aowfi,  
        fafghah\}@clemson.edu}
    \IEEEauthorblockA{\IEEEauthorrefmark{2} Computer Science Department, Vanderbilt University, Nashville, TN, USA \\
        Email: ali.abbasi@vanderbilt.edu}
        
    \IEEEauthorblockA{\IEEEauthorrefmark{3}Air Force Research Laboratory,  Rome, NY, USA \\
	 Emails: \{jonathan.ashdown,kurt.turck\}@us.af.mil}

}

\maketitle

\begin{abstract}
Modulation recognition is a fundamental task in communication systems as the accurate identification of modulation schemes is essential for reliable signal processing, interference mitigation for coexistent communication technologies, and network optimization. Incorporating deep learning (DL) models into modulation recognition has demonstrated promising results in various scenarios. However, conventional DL models often fall short in online dynamic contexts, particularly in class incremental scenarios where new modulation schemes are encountered during online deployment. Retraining these models on all previously seen modulation schemes is not only time-consuming but may also not be feasible due to storage limitations. On the other hand, training solely on new modulation schemes often results in catastrophic forgetting of previously learned classes. This issue renders DL-based modulation recognition models inapplicable in real-world scenarios because the dynamic nature of communication systems necessitate the effective adaptability to new modulation schemes. This paper addresses this challenge by evaluating the performance of multiple Incremental Learning (IL) algorithms in dynamic modulation recognition scenarios, comparing them against conventional DL-based modulation recognition. Our results demonstrate that modulation recognition frameworks based on IL effectively prevent catastrophic forgetting, enabling models to perform robustly in dynamic scenarios.

\end{abstract}

\begin{IEEEkeywords}
Modulation Recognition, Incremental Learning, Deep Learning, Online Learning.
\end{IEEEkeywords}

\section{Introduction}


Modulation recognition is vital in wireless communications for its role in enabling efficient communication, spectrum sharing, and RF fingerprinting. It ensures reliable data transmission, facilitates spectrum allocation in shared environments, and helps identify and secure wireless networks by distinguishing unique RF signatures, making it an indispensable tool in modern wireless systems. In the recent years, deep learning (DL) has led to substantial progress in various fields. Its exceptional ability to derive meaningful information from complex datasets has been particularly influential. This capability has also found effective utilization in communication systems, notably in modulation and signal recognition tasks. In modulation recognition, DL models have demonstrated higher accuracy than traditional methods, particularly in scenarios characterized by significant noise, fading, and impairments\cite{o2017introduction}. This underscores the invaluable role of deep learning in enhancing modulation detection within communication systems, even in challenging operational conditions.

While DL-based techniques have yielded notable outcomes in signal recognition, their practical adoption within communication systems remains limited due to several significant challenges. One of the main obstacles, is the lack of generalization and adaptability of the DL models in varying scenarios, which necessitates continuous retraining and refinement to accommodate new tasks or classes. Furthermore, the requirement for fresh and relevant data adds complexity, leading to computational constraints and time limitations for the retraining process. Addressing these challenges is essential to fully harness the potential of DL methods and integrate them effectively into communication systems.

Most of the previously proposed DL-based solutions for modulation recognition are based on an isolation paradigm, meaning that given a dataset, the DL model is trained, and then tested on the same dataset. This approach does not aim at preserving the acquired knowledge for future learning endeavors and only focuses on the end result. The assumptions used in this paradigm makes it  only suitable for static tasks in closed environments, which is far from the dynamic nature of communication systems with a multitude of varying parameters.

In this paper, we are specifically focusing on a proposing a capable DL-based modulation detection model for a more realistic and dynamic scenario where the receiver encounters signals with new types of modulation (i.e. new classes) as the time grows. In such as scenario, a conventionally trained DL-based modulation detection model has three options, not adjusting the weights for the new classes, retraining on just the new classes, and retraining on all previously seen classes plus the new ones. Option one will simply cause the model to not be able to correctly classify the new classes, and thus, lower its accuracy. Option two will result in catastrophic forgetting\cite{kirkpatrick2017overcoming}, drastic degradation of accuracy on previously learned information when new information is introduced to the model.Lastly, option three is not efficient as it is time and resource consuming and requires memory to keep all the previously seen data, which might not be possible at all depending on the application and circumstances. None of these options are suitable which renders conventional DL-based modulation detection models inapt for the mentioned scenario.

IL is a machine learning approach designed to continually update and enhance a model's knowledge and performance as it encounters new data over time. Its primary purpose is to enable models to adapt to changing environments and evolving datasets without the need for complete retraining and without losing previously learned information. To overcome the aforementioned issues in highly dynamic communication systems with memory-constrained receivers, utilization of incremental learning (IL) for modulation detection has been recently studied \cite{shi2019deep, fan2023c, yang2020modulation}. Although IL has shown promising results in modulation detection in these recent studies, many aspects have still remained unanswered due to limited experiment settings, which are further discussed in detail in the related work section. These limited experiments have resulted in a research gap in the topic of IL-based modulation detection which we aim to fill.

To this end, in this paper we provide an extensive analysis on the performance of IL-based modulation detection methods using multiple different IL algorithms on a large number of modulation schemes, in different continues and dynamic scenarios where new signals with new modulation types are encountered by the receiver. Moreover, we provide a memory study to determine the performance of IL-based modulation recognition models under extremely limited memory budgets.




The rest of this paper is structured as follows: Section II discusses previous papers related to DL-based modulation recognition and incorporation of IL in modulation recognition. Section III formulates the DL-based modulation recognition task in a dynamic class incremental scenario and provides a brief introduction to incremental learning. In Section IV, experiment settings are explained and evaluations are provided. Finally, Section V concludes the paper.
\section{Related Work}

Modulation and wireless signal detection using DL has gained a lot of attention over the last few years. Many different DL architectures have been utilized and tested for modulation classification such as convolutional neural networks (CNNs)\cite{radioml,o2017introduction}, long shortterm
memory (LSTMs) \cite{rajendran2018deep}, Autoencoders\cite{ali2017unsupervised} with competitive accuracy results. While many modulation recognition papers based on conventional DL models have been proposed, some papers have focused on providing non-conventional DL frameworks to improve adaptability and generalization of the DL-based modualtion recognition models. In \cite{owfi2023meta}, the authors proposed a meta learning framework for modulation classification to increase generalization and make the model adaptable to new tasks with a few shots of new data. In \cite{zhou2023semi}, a semi-supervised modulation recognition model was proposed for domain adaptation to environments where the datasets are not well-labeled.

In the more recent works, there are some papers that have applied IL for signal recognition. The authors in \cite{shi2019deep}, briefly discussed the utilization of incremental learning for signal classification in a task incremental learning scenario where the first task contained five types of signals and the second task contained three new types of signals. In \cite{liu2021class}, an IL scheme based on channel separation was proposed for IoT device identification which was evaluated on automatic-dependent surveillance-broadcast data. In \cite{zhou2022electromagnetic}, the authors applied IL for electromagnetic signal classification by using an IL algorithm based on class exemplar selection and a multi-objective linear programming classifier. The authors in \cite{fan2023c}, proposed a complex-valued IL framework for signal recognition, and evaluated their framework in a scenario where the signal recognition model encounters a second group of signal types which were unseen in the initial classes. In \cite{yang2020modulation}, the authors utilized an exemplar-based IL algorithm to classify constellation images of eight signal classes. 

Although all these studies have provided valuable results regarding incorporation of IL in modulation recognition, there are shortcomings in these papers which we aim to improve. Some studies have grouped different modulation schemes into separate tasks in a task incremental scenario\cite{shi2019deep}, which does not correctly reflect how a modulation recognition module at the receiver encounters new modulation schemes in dynamic environments. In some other studies limited number of incremental tasks or classes have been considered \cite{fan2023c,shi2019deep}, which does not effectively demonstrate the efficacy of utilizing IL for modulation detection. Other papers have evaluated limited number baseline and IL algorithms \cite{yang2020modulation}. 

In this paper, we are conducting a thorough examination of modulation detection methods based on the IL framework. We assess their effectiveness by employing a variety of IL algorithms and subjecting them to a comprehensive evaluation involving a large number of modulation schemes. These evaluations are carried out in two dynamic scenarios, replicating situations in which the receiver must contend with new modulation schemes in a class incremental way. Furthermore, we conduct a memory analysis to assess the performance of IL-based modulation recognition models when operating under highly constrained memory sizes.


\section{Methodology}

\subsection{Problem Formulation}

\begin{figure}[htp]
\begin{center} 
\includegraphics[width=0.95\columnwidth]{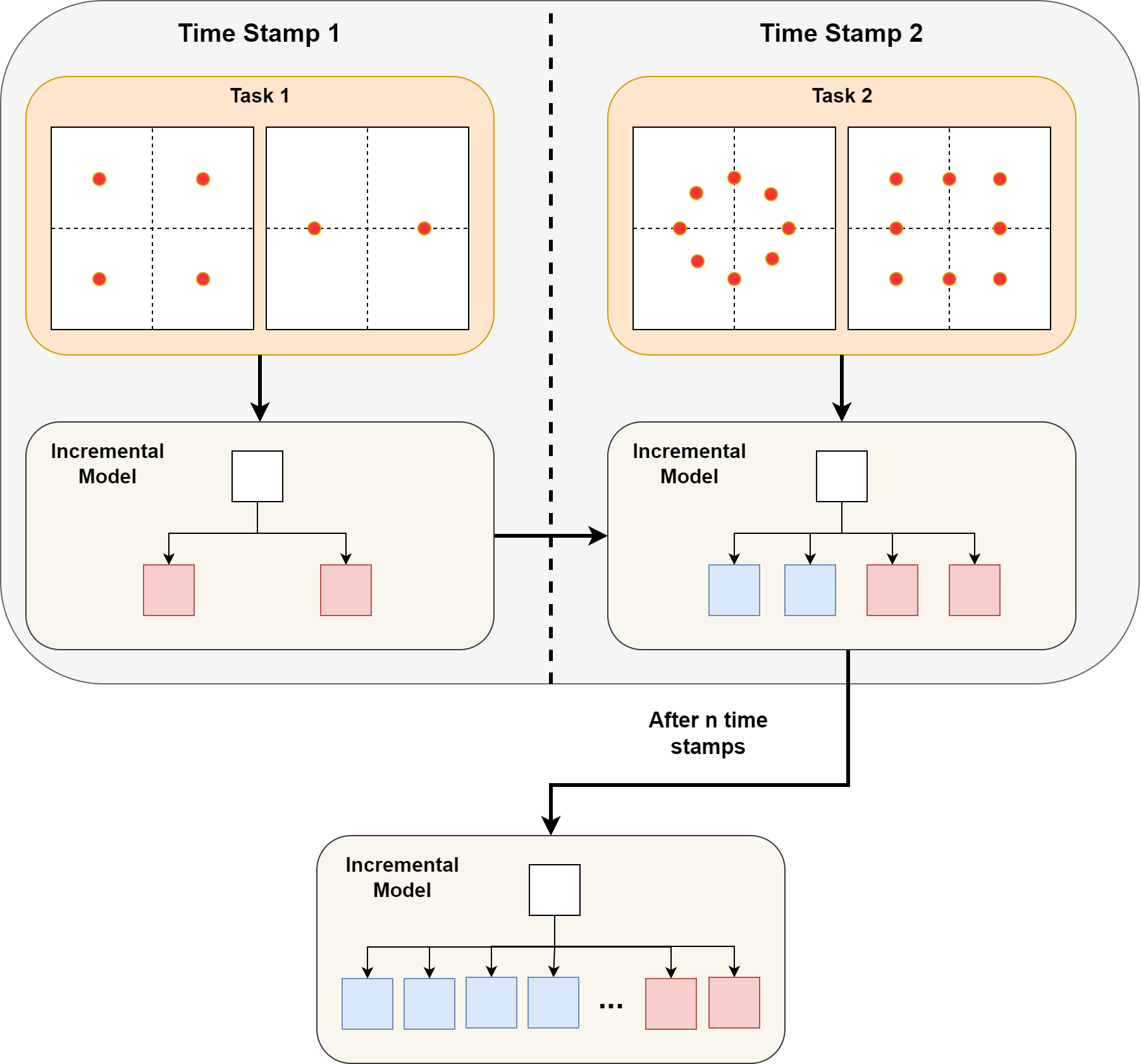}
\caption{Incremental learning modulation classification scenario. Red rectangles and blue rectangles represent classes of newly and previously learned modulation schemes. After each sequence and observing new modulation schemes, the incremental model expands its last layer (referred to as head) to accommodate for the newly encountered modulation schemes.} 
\label{fig:cil_scenario} 
\end{center} 
\end{figure} 

In modulation recognition, the aim is to determine the modulation scheme used in a transmitted signal based on the the received signal. This task can be essentially formulated as a $n$-way classification task where $n$ is the number of possible distinct modulation schemes used at the transmitter.
In a static scenario where the modulation schemes do not change, the problem can be formally written as:

Let $X = \{x_1, x_2, ..., x_m\}$ be a set of I/Q samples of the received signals and $C = \{c_1, c_2, ..., c_n\}$ be a set of modulation schemes. Given a training dataset $D$ with pairs $(x_i,c_j)$:
$$D = \{(x_1,c_1),(x_2,c_2),...,(x_m,c_m)\}$$
Our objective becomes finding a classifier $f: X\xrightarrow{}C$ that accurately determines the modulation scheme used for a given I/Q sample.
However, in realistic scenarios, the receiver does not have access to all modulation classes from the beginning, and in different sequences, new modulation schemes are encountered at the receiver. This scenario, which is illustrated in Figure \ref{fig:cil_scenario}, can be formally written as:

Let $D^{t1}$ be the training dataset available at the receiver at sequence 1 :
$$D^{t1} = \{(x_1^{t1},c_1^{t1}),(x_2^{t1},c_2^{t1}),...,(x_m^{t1},c_m^{t1})\}$$
where each $c_i^{t1}$ belongs to $C^{t1}$, a set of m modulation schemes observed at the first sequence. By training on $D^{t1}$, we find the classifier $f^{t1}$ which is able to classify between the $m$ modulation scheme in $C^{t1}$. In the next sequence, the receiver has access to $f^{t1}$ and a new training dataset $D^{t2}$:
$$D^{t2} = \{(x_1^{t2},c_1^{t2}),(x_2^{t2},c_2^{t2}),...,(x_k^{t2},c_k^{t2})\}$$
where each $c_i^{t2}$ belongs to $C^{t2}$, a set of $k$ modulation schemes observed at the second sequence. Here, our objective is to find classifier $f^{t1+t2}$, that is able to classify between the $m+k$ modulation schemes observed in $C^{t1}$ and $C^{t2}$, given $f^{t1}$ and $D^{t2}$. In another words, the model has to learn to classify the modulation schemes $C^{t2}$ using $D^{t2}$, without the occurrence of catastrophic forgetting on the modulation schemes in $C^{t1}$, as $D^{t1}$ is no longer available to the model at sequence 2. This problem was formulated for 2 sequences but can naturally be extended for $i$ sequences.





\subsection{Incremental learning}

Incremental learning (IL), also refereed to as Continual Learning or Lifelong Learning, is a machine learning paradigm designed to continuously update and expand a model's knowledge base over time, without undermining its previously acquired information. The primary aim of incremental learning is to achieve lifelong learning, where the model evolves to accommodate new information while retaining its proficiency in previously learned tasks\cite{wu2019large,rebuffi2017icarl}. 

A major assumption in IL is unavailability of the whole dataset at a given time, as new classes or tasks are introduced to the model sequentially in time. This assumption represents situations where we are dealing with systems with limited memory and computation resources, such as IoT devices and autonomous drones, very well. Another assumption that is often considered in IL, is having unknown number of classes or tasks, which makes IL models more flexible and especially suitable for applications with evolving and expanding class spaces. Aside from these assumptions, one of the main challenges that IL models have to deal with is the phenomenon known as catastrophic forgetting, where the model's performance on previously learned tasks deteriorates significantly as it acquires new knowledge. 

All these assumptions and challenges essentially translate into two fundamental characteristics of plasticity and stability that need to be carefully balanced to create an effective and adaptable IL model. Plasticity refers to the ability of an IL model to rapidly adapt to new information or data points. A highly plastic model can quickly incorporate new knowledge and adjust its parameters to accommodate changing data distributions or new tasks. Stability, on the other hand, is the capacity of an IL model to retain and consolidate previously acquired knowledge. A stable model should resist forgetting or overwriting existing information when exposed to new data.

The objectives and assumptions of incremental learning align seamlessly with DL-based modulation recognition models designed for dynamic and adaptive environments. Firstly, modulation schemes can fluctuate due to changing channels and hardware impairments, necessitating adaptable modulation recognition models. Additionally, the integration of DL models at transmitters for communication optimization in the future may introduce customized modulation schemes, further emphasizing the need for continuous online learning. Lastly, the inherent constraints of memory and computational resources may restrict access to the entire dataset for retraining at the receiver, underscoring the value of incremental learning for efficient model updates without catastrophic forgetting.


Within the realm of incremental learning, several types of strategies have been proposed to mitigate catastrophic forgetting. Regularization-based approaches employ various regularization techniques to constrain the model's parameters when learning new tasks, reducing the interference with existing knowledge. Parameter isolation techniques isolate and protect specific model components associated with previously learned tasks, allowing new knowledge to be integrated while safeguarding prior expertise. Additionally,Replay-based incremental learning involves storing and periodically replaying samples from previous tasks to maintain their influence on the model and prevent degradation in performance. These samples are often called \textit{exemplars} and can be selected either by chance or a predefined strategy to be better representative of their corresponding tasks/classes while performing in a limited memory budget. 

In class incremental learning with a single incremental head, the model tends to become biased towards the most recent classes. This bias is less severe in task incremental learning with multi-head models, as each task has its own head and the updates for a new task does not update the others. Among the mentioned IL approaches, Replay-based methods can use exemplars to better negate the bias caused in class incremental learning with a single incremental head. To this end, in this paper we utilize three replay-based state-of-the-art IL methods which are briefly introduced below:
\begin{enumerate}
    \item \textbf{iCaRL\cite{rebuffi2017icarl}:} iCaRL proposes to save a memory buffer based on the data similarities to the class prototypes, and performs knowledge distillation between the old and most recent model to prevent catastrophic forgetting. 
    \item  \textbf{BiC\cite{wu2019large}:} In addition to saving randomly chosen exemplars for replaying on the previous tasks, BiC introduces two new parameters to de-bias the unified classifier. These new parameters greatly help the incremental learner to reduce the number of erroneous predictions in favor of the new classes.

    \item  \textbf{LUCIR\cite{hou2019learning}} LUCIR normalizes the feature values right before the final fully-connected layer. Moreover, a margin for each class is imposed on the embedding space such that the features will be more linearly separable, thus, helping the model to prevent interfering with the previously learned embedding space. The memory buffer in LUCIR is comprised based on random selection.
\end{enumerate}



\section{Evaluations}

\subsection{Experiment Settings}


We are using the RADIOML 2018.01A\cite{radioml} signal dataset for our evaluations. This dataset includes over-the-air recordings with synthetic channel effects of 24 types of modulations. These modulations are: OOK, 4ASK, BPSK, QPSK, 8PSK,
16QAM, AM-SSB-SC, AM-DSB-SC, FM, GMSK,
OQPSK , OOK, 4ASK, 8ASK, BPSK, QPSK,
8PSK, 16PSK, 32PSK, 16APSK, 32APSK, 64APSK,
128APSK, 16QAM, 32QAM, 64QAM, 128QAM,
256QAM, AM-SSB-WC, AM-SSB-SC, AM-DSB-WC,
AM-DSB-SC, FM, GMSK, and OQPSK. The dataset includes I/Q samples from -20dB to 30dB with a step size of 2dB. For each class at a specific SNR level there are 4096 I/Q samples of size $2*1024$.

\begin{figure}
\centering
{\includegraphics[width=0.9\columnwidth]{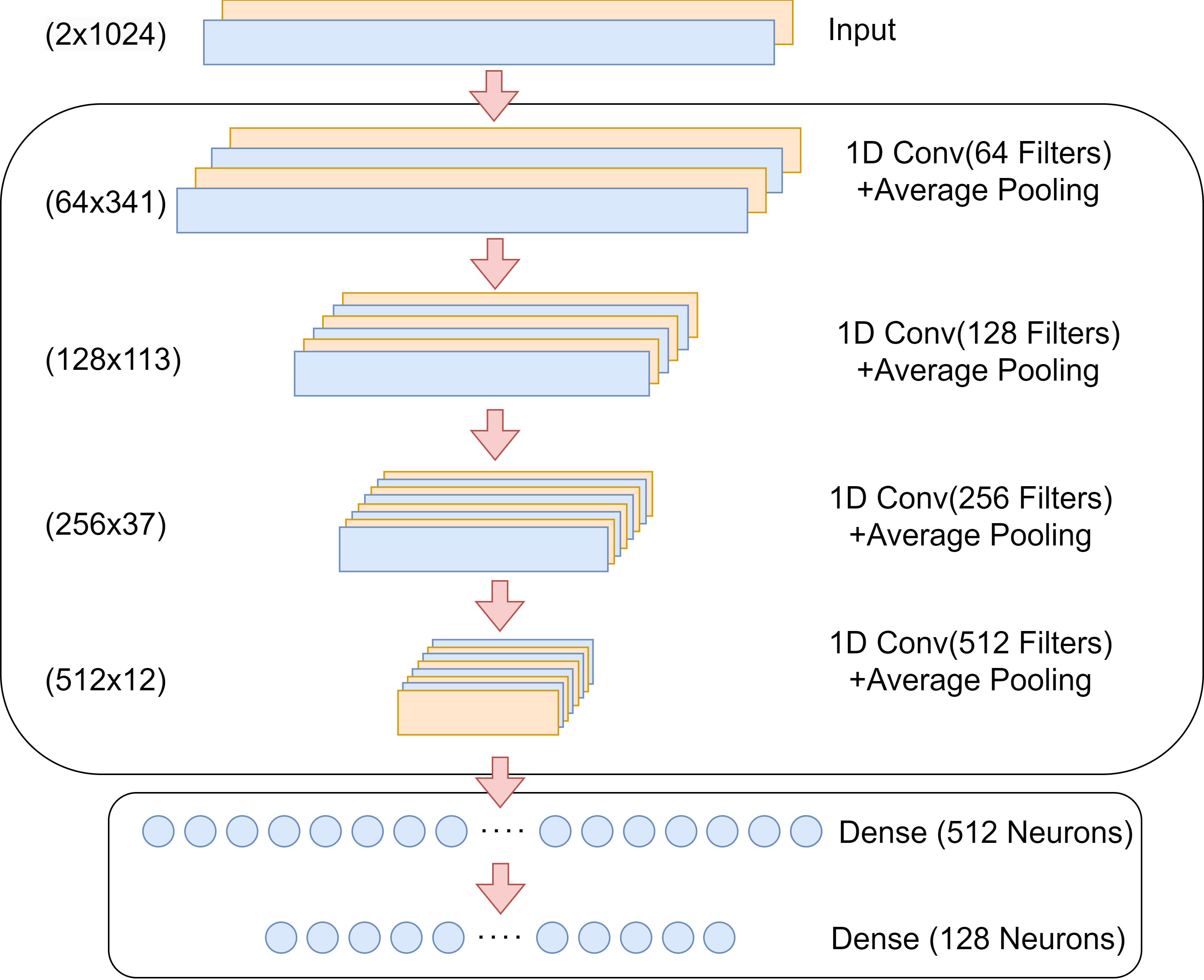}}

\caption{Structure of the backbone neural network. kernel size$=3$, padding$=1$, and ReLU were used for all layers.}
\label{fig:backbone}

\end{figure}

To have a fair comparison among the baseline algorithms, we are using the same neural network structure as the backbone of all the algorithms. The structure of the backbone neural network can be seen in Figure \ref{fig:backbone}. The last layer of the displayed structure is followed by an incremental head with the size of the number of observed classes during training.

For our experiments, we are employing iCaRL, BiC, and LUCIR, which we introduced previously, for our IL-based modulation recognition models. Moreover, we are using two non-IL baselines for our experiments so that we can better evaluate the performance of modulation recognition based on the IL framework. The two baselines are as follows:

\begin{itemize}
    \item Conventional DL model: During each timestamp, the backbone model updates itself by training on the available dataset at that timestamp.
    \item Joint Learning: Training the backbone model with the dataset of all observed modulation schemes at each timestamp. Joint learning essentially represents the ideal case of not having any retraining or memory limitations as we are assuming the dataset of all previously modulation schemes are available for retraining at any point. Thus, this baseline provides us with an upper bound on the achievable accuracy for the other methods.
    
\end{itemize}

\subsection{Accuracy experiment}

\begin{figure}
\subfloat[][SNR=0]{\includegraphics[width=0.49\columnwidth]{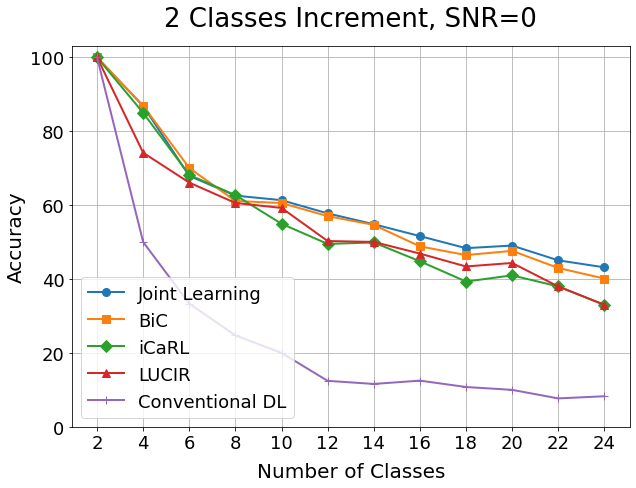}}
 \subfloat[][SNR=20]{\includegraphics[width=0.49\columnwidth]{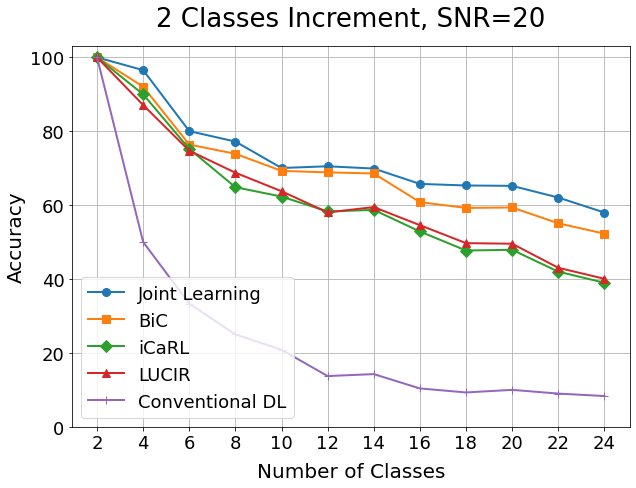}}

\caption{Accuracy of modulation recognition models in 2 classes increment scenario (12 tasks in total). Figure (a) and (b) depict results on SNR=0 and SNR=20 respectively.}
\label{fig:2cic_acc}
\end{figure}

Figure \ref{fig:2cic_acc} presents the accuracy of IL-based modulation recognition models and the baseline models in a 2 class increment scenario in high SNR and low SNR cases. All three IL-based modulation recognition models are using 2000 exemplars for this experiment. As it can be seen from the figure, the conventional modulation recognition model suffers from catastrophic forgetting and fails to preserve the learned information from previously observed classes. The performance of IL-based models, specially BIC, on the other hand are very close to the upperbound joint learning baseline. This illustrates with an incremental learning framework, we can effectively adapt to new classes while preserving previously learned information and without the need of storing all the previous datasets. 

\begin{figure}
\subfloat[][SNR=0]{\includegraphics[width=0.49\columnwidth]{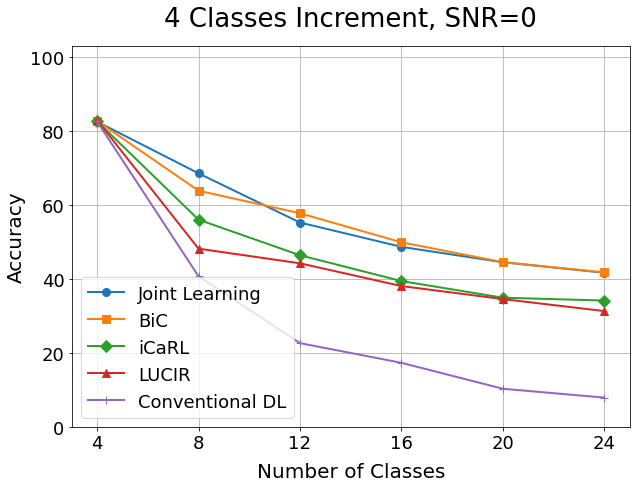}}
 \subfloat[][SNR=20]{\includegraphics[width=0.49\columnwidth]{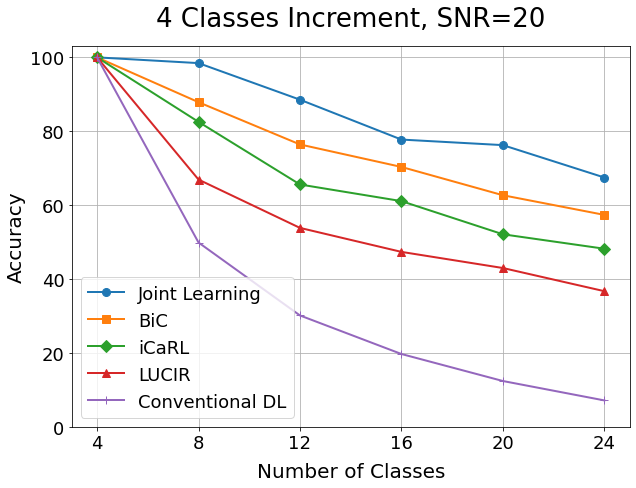}}

\caption{Accuracy of modulation recognition models in 4 classes increment scenario (6 tasks in total). Figure (a) and (b) depict results on SNR=0 and SNR=20 respectively.}
\label{fig:4cic_acc}

\end{figure}




To further investigate the performance of IL-based modulation recognition models, we also considered a case of 4 class increment with 6 sequential tasks in total. Figure \ref{fig:4cic_acc} provides accuracy of compared models in the 4 class increment scenario. As it can be seen in the figure, we can again clearly observe catastrophic forgetting in the conventional DL model. For the IL-based modulation recognition models, we can observe that catastrophic forgetting is effectively negated in BiC, specifically in SNR=0, where BiC has a near-to-ideal accuracy of joint learning. iCaRL and LUCIR also have a good performance in SNR=0 with respect to the conventional DL model and the joint learning model. In SNR=20, we can see that LUCIR's becomes less effective in comparison to the other IL models. It should be noted that there are 4096 samples per modulation and SNR level in the dataset we are using. This means there are a total of $2000*24=48000$ samples in total that the Joint Learning model has access to at the last stage of the scenario, whereas all the IL-based modulation recognition models are only storing 2000 samples in total in their memory while performing nearly as good as the upperbound,specially BiC, illustrating the effectiveness and efficiency of these models for dynamic modulation recognition.

\subsection{Memory experiment}




\begin{figure}
\centering
{\includegraphics[width=0.8\columnwidth]{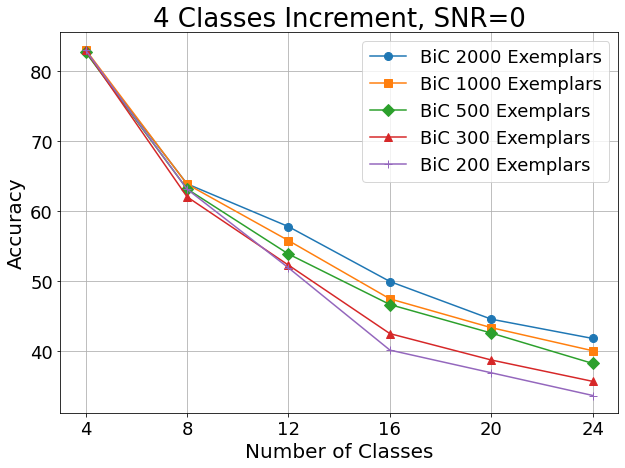}}

\caption{Accuracy of IL-based modulation recognition model based on BiC in a 4 class increment scenario, with different memory budgets.}
\label{fig:memory}

\end{figure}

In the previous section, we compared the accuracy of IL-based modulation recognition models in a dynamic class incremental scenario against conventional DL modulation recognition and an upper bound baseline which was Joint learning. The results showed us that by using an IL-based framework, we can effectively negate catastrophic forgetting for the previously observed modulation schemes. Among the compared IL algortihms used for modulation recognition, BiC provided the best results. As the BiC takes the approach of selecting and storing small proportion of samples (referred to as exemplars) from the whole dataset of the previously observed classes, we are interested in knowing the trade-off between the memory usage and accuracy of IL-based modulation recognition model using BiC.

Figure \ref{fig:memory} provides accuracy of the IL-based modulation recognition model using using BiC in five different cases of exemplar memory size, from just 200 samples to 2000. This experiment is done using SNR=0 data and in a 4 class incremental scenario with 24 classes in total. For 200, 300, 500, 1000, and 2000 available number of exemplars, the accuracy of the recognition model in the last stage is 33.64, 35.66, 38.22, 40.04 and 41.80 percent, respectively. As observed in Figure \ref{fig:4cic_acc} (a), the last stage accuracy of the joint learning model (upperbound) and the conventional DL model in the same scenario was 41.78 and 9.83 percent, respectively. These results show that even in extremely limited memory cases where the memory budget allows for storing 200 samples, utilizing IL algorithms such as BiC can still effectively prevent catastrophic forgetting to a good extent and thus, significantly improve modulation recognition accuracy in dynamic online scenarios

\section{Conclusion}

In this paper, we demonstrated the efficacy of incremental learning-based framework for wireless signal modulation recognition in dynamic online scenarios, where the receiver encounter new modulation schemes over time. Our results illustrated that while conventional DL-based modulation recognition models are not capable of keeping information of past observed modulation schemes, IL-based modulation recognition models are capable of both keeping past information and also adapting to new modulation schemes. Among the IL algorithms that we used for IL-based modulation recognition, BiC provided the best results, with the near-to-ideal accuracy in SNR=0 cases. Moreover, in another experiment  done to test the effectiveness of IL-based modulation recognition models with differed memory sizes, we showed that even in extremely limited memory cases where only 200 samples can be stored by the model, we can still significantly improve accuracy in dynamic scenarios by employing the IL framework for modulation recognition.




\bibliographystyle{IEEEtran}
\bibliography{main}

\end{document}